# Search for supernova-produced $^{60}$Fe in a marine sediment

C. Fitoussi[1], G. M. Raisbeck[1], K. Knie[2,5], G. Korschinek[2], T. Faestermann[2], S. Goriely[3], D. Lunney[1], M. Poutivtsev[2], G. Rugel[2], C. Waelbroeck[4], A. Wallner[5]

[1] Centre de Spectrométrie Nucléaire et de Spectrométrie de Masse (CSNSM) IN2P3/CNRS, Campus d'Orsay, Bât 108, 91405 Orsay, France
[2] Technische Universität München, Fakultät für Physik, James-Franck-Straße 1, 85747 Garching, Germany
[3] Institut d'Astronomie et d'Astrophysique, C.P.226, Université Libre de Bruxelles, bd. Du Triomphe, B-1050 Brussels, Belgium
[4] Laboratoire des Sciences du Climat et de l'Environnement (LSCE), Domaine du CNRS, Bât 12, 91198 Gif-sur-Yvette, France
[5] VERA Laboratory, Fakultät für Physik, Universität Wien, Währinger Str. 17, A-1090 Vienna, Austria

**Abstract**

An $^{60}$Fe peak in a deep-sea FeMn crust has been interpreted as due to the signature left by the ejecta of a supernova explosion close to the solar system $2.8 \pm 0.4$ Myr ago [Knie et al., Phys. Rev. Lett. **93**, 171103 (2004)]. To confirm this interpretation with better time resolution and obtain a more direct flux estimate, we measured $^{60}$Fe concentrations along a dated marine sediment. We find no $^{60}$Fe peak at the expected level from 1.7 to 3.2 Myr ago. However, applying the same chemistry used for the sediment, we confirm the $^{60}$Fe signal in the FeMn crust. The cause of the discrepancy is discussed.

Some recent studies suggest that one or more supernovae (SN) could have exploded near the solar system in the last ten million years [1,2]. Ellis et al. [3] discussed possible isotopes that might be used to look for traces of such a nearby SN in geological reservoirs. Indeed, an indication for an $^{60}$Fe ($t_{1/2}$ = 1.49 Myr) excess has been observed [4] and later on definitely been measured in a

FeMn crust [5]. This excess was interpreted as the signature left by a nearby SN explosion 2.8 ± 0.4 Myr ago. In this letter, we present results of a concerted effort to confirm the $^{60}$Fe result of [5], including a different chemical treatment of the same FeMn crust, as well as using a different geological reservoir: marine sediment. The faster accumulation rate of sediments potentially allows a refinement of the time resolution, as well as a more direct estimate of the flux.

The sediment was sampled from an ODP core, Leg 162, site 985 (66°56.5' N, 6°27.0' W) in the North Atlantic [6]. The average sedimentation rate is 3 cm/kyr, with an in-situ density of ~1.6 g/cm$^3$ and an average iron concentration in the authigenic phase of ~0.5 wt%. A continuous sequence of 30 cm long samples was taken, corresponding to time intervals of 10 to 15 kyr each. Details of the dating of the sediment core - using paleomagnetic information - are given elsewhere [7].

FeMn crusts are authigenic objects, that is, composed of species precipitated and adsorbed from the soluble phase of the ocean. The deep sea sediment chosen in the present study is characterized by an accumulation rate about three orders of magnitude higher than that of FeMn crusts, and therefore allows a much better time resolution. However, sediments are composed of a larger variety of geochemical phases, particularly the alumino-silicate phase which contains most of the stable Fe of the sediment. Since the iron in this alumino-silicate phase is not equilibrated with the soluble $^{60}$Fe in the ocean, its inclusion would lower the $^{60}$Fe/Fe ratio compared to the authigenic phase. Therefore, we used a chemical procedure to isolate the authigenic fraction. The chemical procedure has been detailed elsewhere [7,8]. Briefly, the sediment was leached with a solution of hydroxylamine hydrochloride in an acetic acid medium, iron was isolated by ion-exchange chromatography and finally precipitated in the form of iron hydroxide. The AMS measurements were carried out at the facility of Munich [9]. In order to be able to measure ultra-low concentrations of that many samples, some samples were combined in groups of two or four.

To verify that our chemical procedure was indeed dissolving the phase containing the $^{60}$Fe seen in the crust, we tested this chemistry on samples from a drill hole of the same crust used by [5], 237KD from the cruise VA 13/2 [10]. Although our statistics are more limited, we observed again an $^{60}$Fe/Fe peak, consistent with that found by [5] (see Fig. 1).

The $^{60}$Fe signal in the FeMn crust was observed over a depth interval corresponding to about 800,000 years [5]. However, the estimated deposition time of a SN ejecta on Earth is likely to be of order ~10,000 years [11], and the residence time of iron in the ocean is much smaller than this value. Because it can be sampled at much higher time resolution, we expected the $^{60}$Fe/Fe signal in the sediment to be much higher than in the crust. While iron can be reduced and become mobile in some sedimentary systems, this effect is generally limited to a few tens of cm below the sediment surface. Because our samples are 30 cm in length, we do not expect such an effect, if present, would significantly distort a signal of this order. On the basis of the measured $^{60}$Fe/Fe in the crust and an estimated uptake factor, Knie et al. [5] inferred an $^{60}$Fe fluence of $2 \times 10^9$ at/cm$^2$ in the interstellar medium. The latter corresponds to an $^{60}$Fe fluence on Earth of $1.4 \times 10^8$ at/cm$^2$ when corrected for radioactive decay. Using this fluence, a deposition time of 10,000 years, and the characteristics of the marine sediment core mentioned above (1.6 g/cm$^3$ density, [Fe]$_{authigenic}$ = 0.5 wt%, sedimentation rate of 3 cm/kyr), we can calculate that the expected $^{60}$Fe/Fe ratio in the authigenic phase of the sediment is $5 \times 10^{-14}$ (see Fig. 2). If the sampling or the effective measuring interval is longer than 10,000 years, then the expected signal will be proportionally reduced. This is the case for the time intervals corresponding to samples which were grouped by two or four for the AMS measurements, as mentioned above.

As can be seen on Fig. 2, the measured $^{60}$Fe/Fe ratio along the sediment core from 2.4 to 3.2 Myr is much lower than the expected value. There are at least five possible explanations: (i) The dating of the $^{60}$Fe signal in the crust is in error (ii) The deposition time of $^{60}$Fe is much longer than the simple expectation for a SN. (iii) The interstellar fluence derived by [5] is overestimated due to an error in the uptake efficiency. (iv) Our sediment core has not recorded the expected global signal. (v) The excess of $^{60}$Fe observed in the FeMn crust is from another source which we have not identified.

First, the dating of the crust reported in [5] was taken from $^{10}$Be measurements on the same crust [13], but with material from a drill hole separated by a distance of at least 20 cm from that used for the $^{60}$Fe measurements. For this reason, the $^{10}$Be dating has been repeated in another drill hole right next to that one where the $^{60}$Fe signal was found. Twelve layers of 1 mm were milled off and dissolved in aqua regia after adding about 2 mg stable beryllium carrier. Sample preparation was carried out as described in reference [14]. The $^{10}$Be/$^{9}$Be ratios have been measured by AMS at the VERA laboratory in Vienna [15], a dedicated AMS facility based on a smaller tandem accelerator (3 MV terminal voltage) which allows more precise measurements. The results are depicted in Fig. 3. Except for the samples near the surface, one can see a nearly exponential decrease of the $^{10}$Be concentration which indicates a constant growth rate in the time span around the $^{60}$Fe signal. Near the surface the $^{10}$Be data suggest either rapid diffusion or a much higher growth rate in recent time. The latter explanation was already postulated by [13] from $^{230}$Th measurements.

To determine the age $T_n$ of a layer of $n$ to $n+1$ mm depth, we use the relation $T_n = \ln(C_0/C_n) \tau_{10}$, with $C_n$ as the $^{10}$Be concentration of the layer from $n$ to $n+1$ mm depth and $\tau_{10}$ as the mean life time of $^{10}$Be. Using a half-life of 1.51 Myr for $t_{1/2}(^{10}$Be$)$, and neglecting the outer 2 mm, the depth interval of the $^{60}$Fe signal (6-8 mm) corresponds to a time span of 1.96 Myr to 2.86 Myr. This compares with the interval 2.4-3.2 Myr cited by [5] based on the dating of [13] but which included the first 2 mm, and used an average between growth rates deduced from $^{10}$Be concentrations and $^{10}$Be/$^{9}$Be ratios.

A second source of uncertainty on the dating, which was not taken into account in the estimate of the age interval in [5], is from the half-life of $^{10}$Be on which there is still a debate in the literature, with values ranging from 1.34 to 1.51 Myr [16-18], the latest reported value being 1.36 ± 0.07 Myr [18]. Using a low value of 1.34 Myr, the time interval found above for the FeMn crust becomes 1.74 to 2.54 Myr. We thus decided to extend the measurements of the $^{60}$Fe/Fe ratios along the sediment core from 1.7 to 2.4 Myr.

As can be seen in Fig. 4, the $^{60}$Fe/Fe signal observed over this time interval is also much smaller than the predicted value. Around t ~2.25 Myr samples were measured for longer times, because we initially had a spurious indication of a peak in this region. However, the upper limits for these measurements are comparable to those in the other time range.

We thus have no evidence for an $^{60}$Fe signal at the levels expected based on the local interstellar $^{60}$Fe fluence given by Knie et al. [5], and the assumptions described above. One of the most important of these is the assumed deposition time of 10 kyr. The $^{60}$Fe signal observed by [5] corresponded to a time interval of ~800 kyr. However, we assumed that this was due to the inherent time resolution associated with the growth and sampling of the FeMn crust. It is interesting to examine what limits we can set for a signal >>10 kyr. To do this, we show in Fig. 5 the running means calculated by combining our data in intervals of ~400 and 800 kyr respectively. There are two ways of interpreting these figures. If we neglect the structure in the curves, and assume all the events are background, we get an average value for $^{60}$Fe/Fe of 2.3 ± 0.3 x10$^{-16}$, as shown by the horizontal dotted lines in Fig. 5a. Within 2 σ, this is consistent with all the data in Fig. 5a, and is almost exactly equal to the background signal found by Knie et al. [5], with the same AMS setup, in the crust beyond the peak region. Using the general criteria of a detection limit being 3 times the standard deviation of background, this would imply a detection limit of 0.9x10$^{-16}$, which is ~7 times lower than the predicted value using the fluence of [5]. If we treat the

400 kyr running means by the same criteria, the detection limit is ~15 times lower than the predicted value.

Alternatively, we can note that the lowest observed signal ($1.0 \pm 0.4 \times 10^{-16}$) in Fig. 5b, at ~1.9 Myr, is significantly less than the expected background. If we consider this as the real background for our measurements, we find signals of marginal significance in the 400 kyr running mean of $2.6 \pm 0.8 \times 10^{-16}$ centered at ~2.4 Myr and $2.9 \pm 1.7 \times 10^{-16}$ at 2.65 Myr. While these are about 5 times lower than the predicted value, this might be accommodated by the uncertainty in the uptake factor of [5] (this factor is in fact very poorly constrained), or by our assumption that the $^{60}$Fe fluence at the location of our sediment is quantitatively representative of the average global value. While statistically valid, we hesitate this latter interpretation of our results because it requires that the background level was rigorously constant for all our measurements (carried out over several months). But the sources of the background are not fully identified and only speculations are possible. Thus, for the moment we prefer to interpret our results only as upper limits.

What would be the implications of a $^{60}$Fe signal lasting >>10 kyr? It is unlikely that such a signal could originate from a fast (ie pressure sufficient to overcome the solar wind) SN shock wave traversing the solar system. One might imagine, however, such a signal coming from the encounter of the solar system with a locally enhanced concentration of SN ejecta from a shock wave that had slowed down, or even stopped, relative to the local interstellar reference frame (LSR). If we assume, for example, a velocity of ~15 km/s for the solar system relative to the LSR, then the size of a feature traversed in 400 kyr is ~6 pc. This is roughly the size of the local interstellar cloud in which the solar system is currently embedded [19]. It is believed that there are many such warm clouds within the Local Bubble, a region of hot, low density gas in which the solar system has been travelling for several million years. While there is still considerable discussion on the exact formation mechanism of the Local Bubble (see for example [20] and refs within), most of these involve SN explosions in one way or the other. Thus it seems quite plausible that some of the clouds in the bubble contain relatively fresh SN ejecta, and that the solar system encountered such a cloud ~2.5 Myr ago.

If the SN ejecta did not have sufficient pressure to overcome the solar wind, then the $^{60}$Fe could not have entered the solar cavity in ionized form, but would instead have to have been in the form of neutral atoms or condensed material with a relatively large size of >0.1 micron (smaller particles are believed to be excluded due to their charge [21]), see also discussion in [4]. Since most of the Fe in the warm clouds is believed to be in the condensed form [20], this latter scenario is perhaps not unreasonable. A related question is how such Fe could have been incorporated into the FeMn crust? Either it would require that the particles containing the Fe were dissolved in the ocean, and then incorporated in the authigenic phase, or that they accumulated as solid particles in the crust, but were dissolved during the extraction process, including that used in the present study. Regarding this fraction, we would expect to find a corresponding signal in our sediment.

In fact, Basu et al. [22] have recently argued that the $^{60}$Fe signal observed by [5] results from incorporation of "micrometeorites" in the crust (they do not make it clear whether they are referring to actual micrometeorites i.e. objects which have been exposed in interplanetary space as small objects, or atmospheric ablation products of much larger "classical" iron meteorites). The $^{60}$Fe is assumed to have come from the interaction of galactic cosmic rays with the Ni in these objects. They estimate the $^{60}$Fe concentration from that found in iron meteorites, typically $8 \times 10^8$ atoms/g Ni [23]. This can be compared to the concentration found by [5] in the crust (~$1 \times 10^7$ atoms/g crust when corrected for decay). This would require that the crust be made up of ~1.3 wt% pure Ni extraterrestrial spherules, (and a correspondingly larger fraction for more realistic Ni concentrations). Taking the case of a homogeneous distribution of small particles of this kind, and knowing that the total Ni concentration in the crust, including the layer containing the $^{60}$Fe, is ~0.5 wt% [24], this hypothesis is clearly untenable. The case for large particles accounting for the $^{60}$Fe in the crust is also unrealistic since the signal was observed every time it was attempted to be measured: on successive layers 6-7 mm and 7-8 mm, as well as on the 6-8 mm layer [5], and after drilling a new hole in this crust (this work, Fig. 1). We also do not understand the argument of Basu et al., who claim that a single micrometeorite of 500 µm diameter having the $^{60}$Fe concentration in the meteorite Dermbach (ie ~0.58 mg if pure Ni containing $4.6 \times 10^5$ atoms of $^{60}$Fe) can account for the $^{60}$Fe measured in the crust [5], that is, $3.9 \times 10^6$ atoms $^{60}$Fe knowing that the section drilled in the crust by [5] has a diameter of 13 mm.

In summary, our results appear to be inconsistent with the traversal of the solar system by a young SN shock wave having the $^{60}$Fe fluence estimated by [5]. Our upper limits are also lower than the predicted value for longer duration signals such as those that might result from the interaction of the solar system with SN ejecta that has greatly slowed down or come to rest with respect to the LSR. The most optimistic interpretation of our results would allow one (or two) signals of ~400 kyr duration and a fluence ~5 times less that estimated by [5]. The discrepancy could possibly be due to the fact that our sediment has not quantitatively registered the average global signal. Additional experiments with other sediment cores would help to clarify this possibility. More probably, however, is an overestimate of the $^{60}$Fe fluence in [5]. Some of the present authors are involved in experiments to remeasure $^{53}$Mn in Antarctic ice which might help to better constrain the $^{53}$Mn uptake factor on which the $^{60}$Fe fluence is based.

ACKNOWLEDGMENTS

We wish to thank the Ocean Drilling Program for coring operation of Leg 162 and Walter Hale for his help in providing the sediment samples. This work was partially supported by the EC under contract number HPRI-CT-2001-50034 and by France's IN2P3 program DESTIN (PICS 2890).

**FIG. 1.** Confirmation of the $^{60}$Fe signal of Knie et al. [5] in the same FeMn crust using the chemical leaching procedure developed for the analysis of the sediment.

**FIG. 2.** Measured $^{60}$Fe/Fe (note log scale) along the sediment core during the time span originally defined by the $^{60}$Fe signal observed in the FeMn crust. The vertical error bars (68.3 % confidence level) were calculated using counting statistics appropriate for small numbers [12]. The width and amplitude of the expected SN signal (shown at 2.8 Myr) was calculated assuming samples were measured individually or combined by two or four (see text), as indicated by the horizontal error bars on the experimental data. The horizontal line at $2.4 \times 10^{-16}$ is the background level given by Knie et al. [5].

**FIG. 3.** $^{10}$Be concentration in atoms/g crust versus the depth of the layer. The horizontal error bars indicate the depth interval covered by the respective layer. The line is an exponential fit for the region 2 – 10 mm. The slope corresponds to a growth rate in this time span of 2.14 mm/Myr (with $t_{1/2}(^{10}\text{Be}) = 1.51$ Myr).

**FIG. 4.** $^{60}$Fe/Fe along the sediment core during 1.7 to 2.4 Myr ago (showed as in Fig. 2).

**FIG. 5.** Data in Figs 2 and 4 plotted as (a) 800 kyr and (b) 400 kyr running means and standard deviation. The dashed histograms represent the expected ratio in our sediment calculated from the fluence given by Knie et al. [5]. Horizontal dotted lines in (a) correspond to ± one standard deviation limits of the whole data set, and in (b) correspond to the lowest significant signal measured during these experiments (see text).

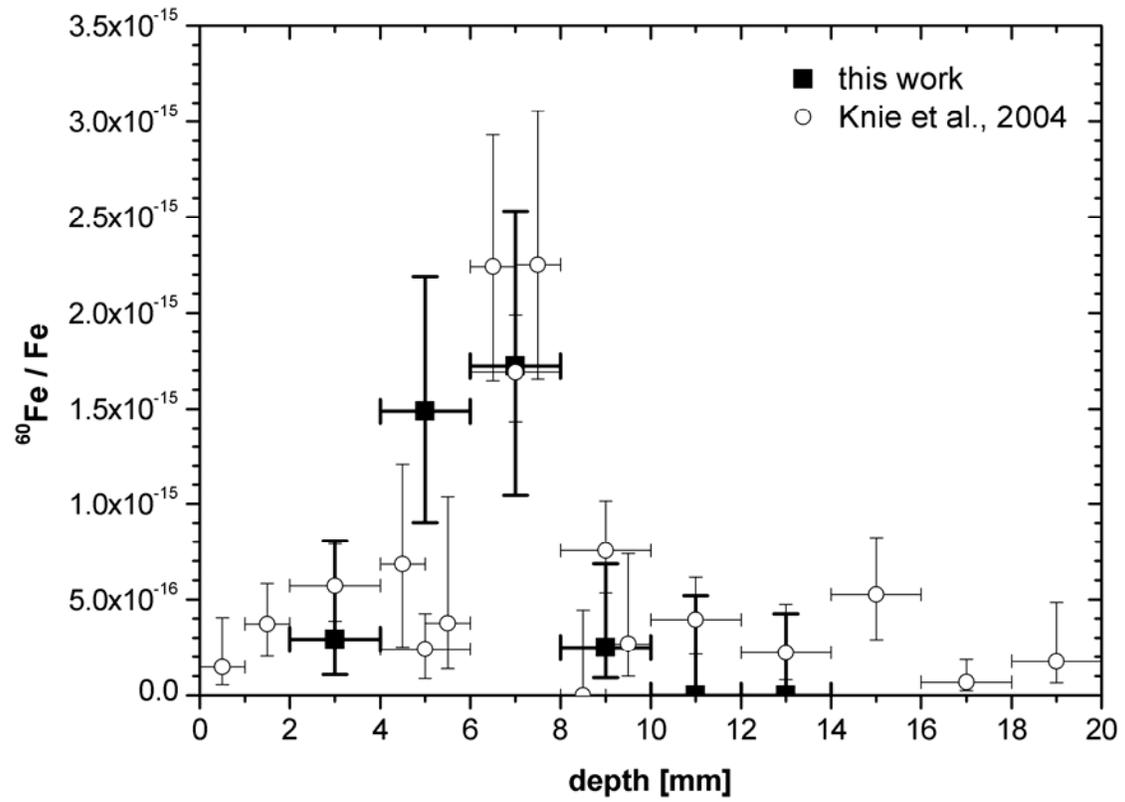

**FIG. 1.**

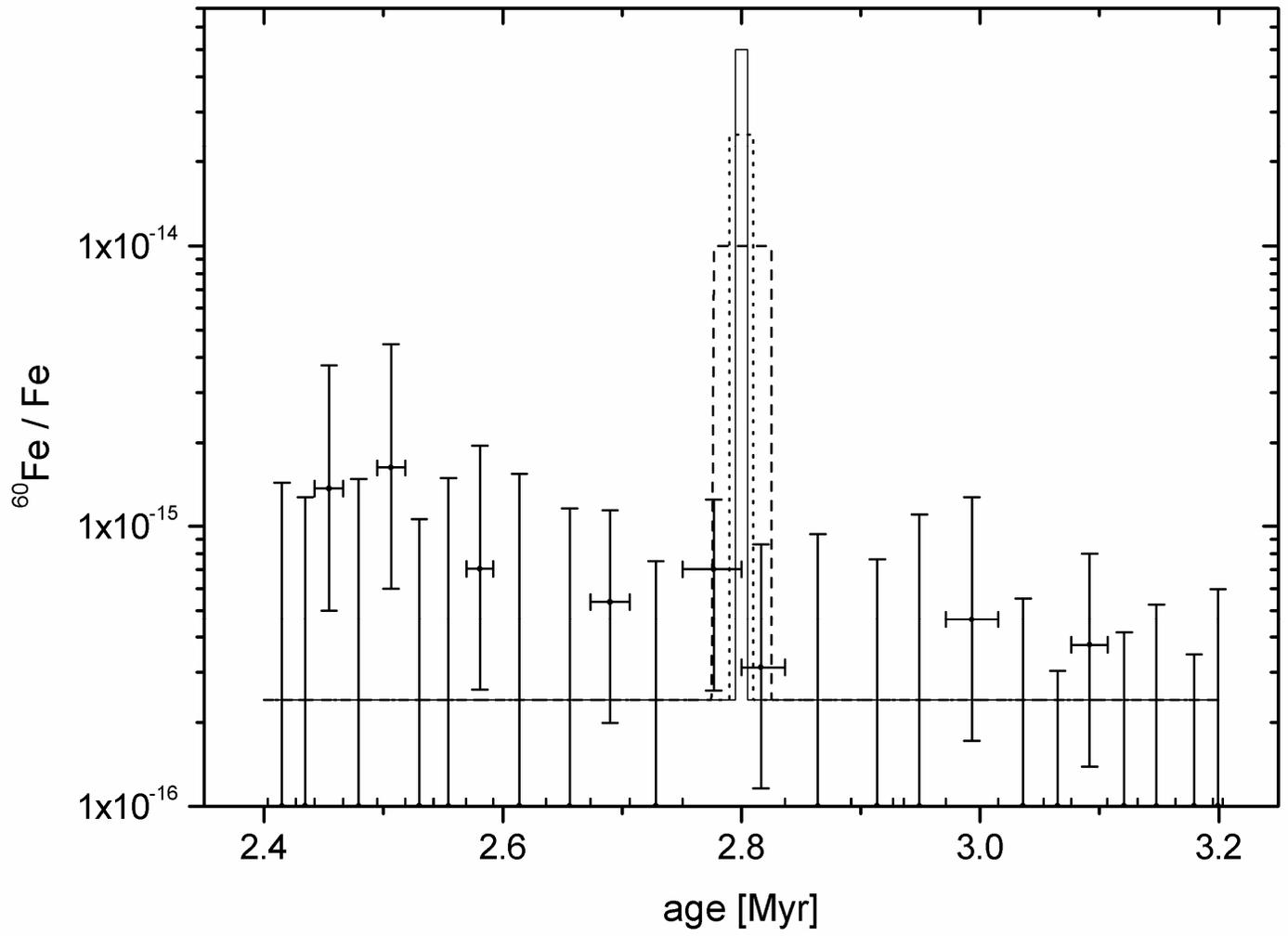

**FIG. 2.**

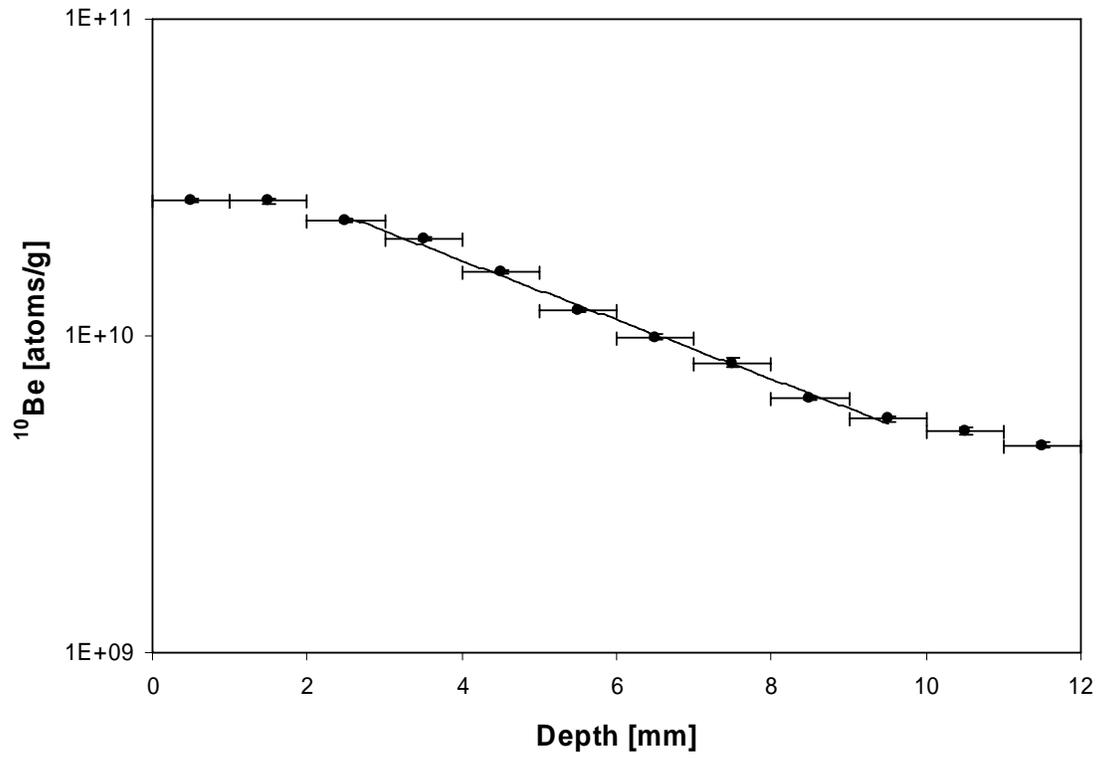

**FIG. 3.**

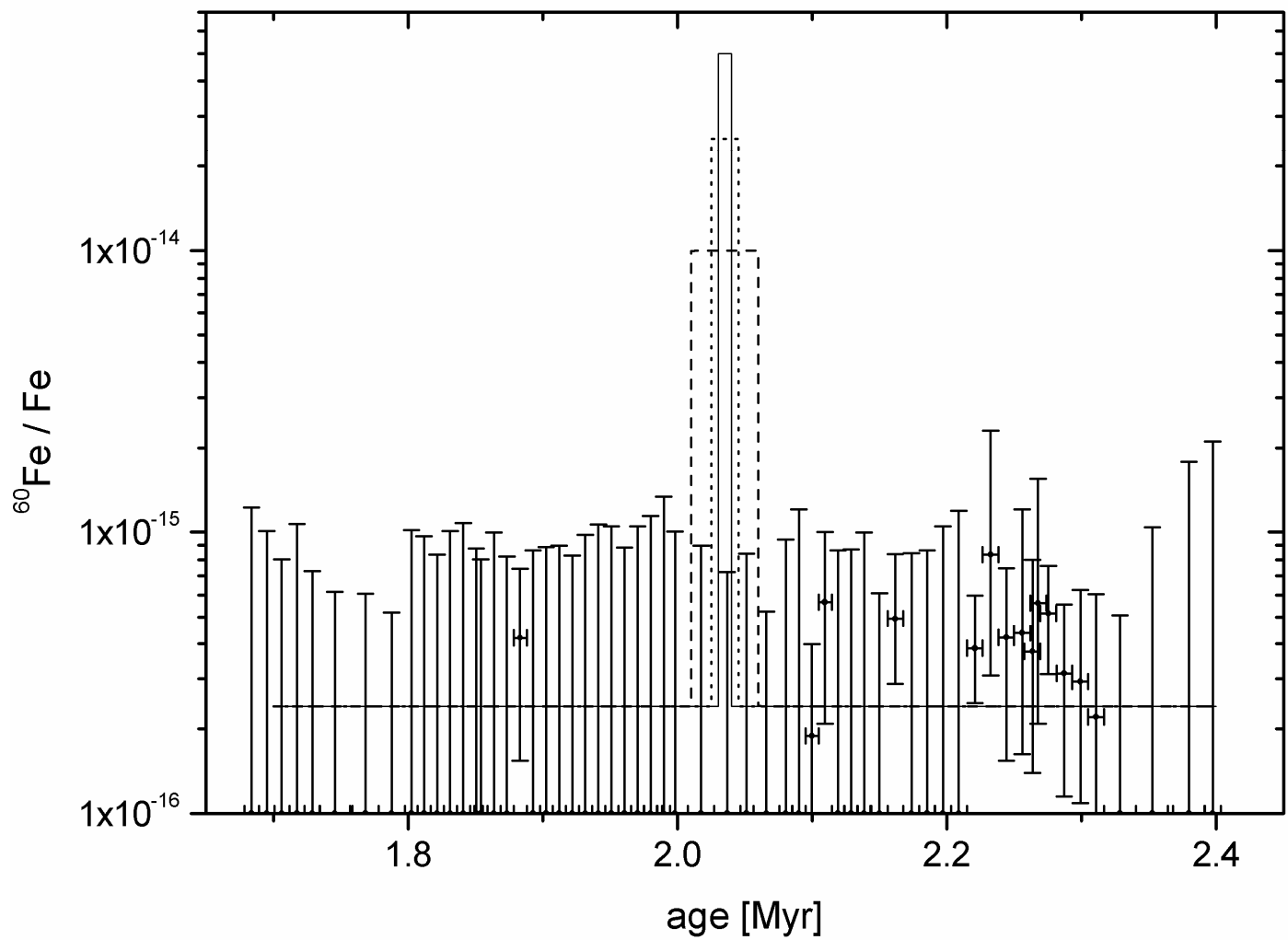

**FIG. 4.**

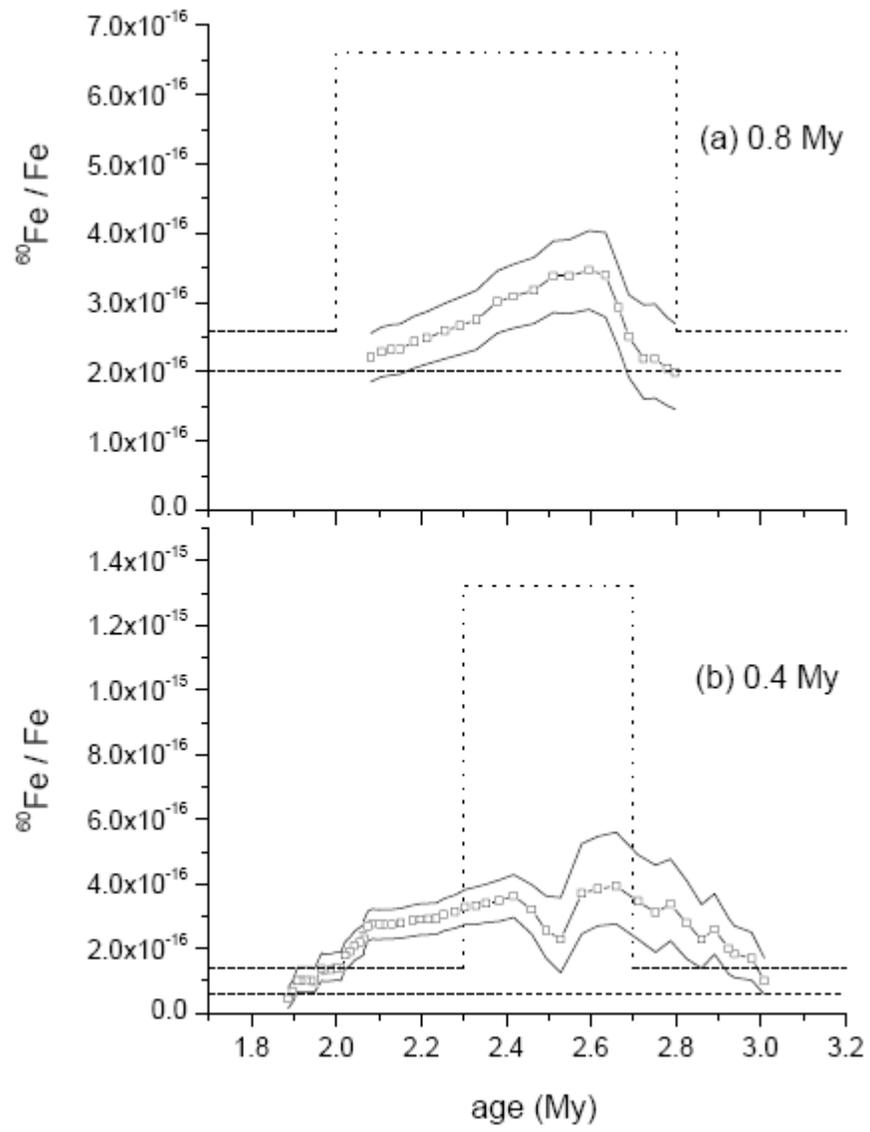

**FIG. 5.**